\documentclass[checkin,showpacs,aps,prl]{revtex4}
\usepackage{slashbox}
\usepackage{threeparttable}
\usepackage{amsmath}
\usepackage{dcolumn}
\usepackage{multirow}
\usepackage{booktabs}
\usepackage{graphicx,color}
\usepackage{graphicx}
\usepackage{amsmath}
\newcolumntype{d}[1]{D{.}{.}{#1}}
\newcommand{\ba}{\begin{eqnarray}}
\newcommand{\ea}{\end{eqnarray}}

\newcommand{\be}{\begin{equation}}
\newcommand{\ee}{\end{equation}}

\newcommand{\et}{{\it et al. }}

\def\prl{{ Phys. Rev. Lett. }}
\def\apb{{ Appl. Phys. B }}

\def\apl{{ Appl. Phys. Lett. }}
\def\prb{{ Phys. Rev. B }}

\def\jap{{J. Appl. Phys. }}
\def\njp{{N. J. Phys. }}
\def\np{{Nature Phys. }}
\def\jpcm{{J. Phys.: Condens. Matter }}

\begin{document}


\title{Resolving photon-shortage mystery in femtosecond magnetism}

\author{M. S. Si}

 \affiliation{Department of Physics, Indiana State University, Terre
  Haute, Indiana 47809, USA\\
  Key Laboratory of Magnetism and Magnetic Materials of the Ministry of Education,
  Lanzhou University, Lanzhou 73000, China}

\author{G. P. Zhang$^{*}$}
  \affiliation{Department of Physics, Indiana State University, Terre
  Haute, Indiana 47809, USA}

\date{\today}
\begin{abstract}
{
For nearly a decade, it has been a mystery why the small average 
number of photons absorbed per atom from an ultrashort laser pulse is able to
induce a strong magnetization within a few hundred femtoseconds. Here we
resolve this mystery by directly computing the number of photons per atom
layer by layer as the light wave propagates inside the sample. We find that
for all the 24 experiments considered here, each atom has more than one
photon. The so-called photon shortage does not exist. By plotting the
relative demagnetization change versus the number of photons absorbed per
atom, we show that depending on the experimental condition, 0.1 photon can
induce about 4\% to 72\% spin moment change. Our perturbation theory reveals
that the demagnetization depends linearly on the amplitude of laser field.
In addition, we find that the transition frequency of a sample may also play a
role in magnetization processes. As far as the intensity is not zero, the
intensity of the laser field only affects the matching range of the transition
frequencies, but not whether the demagnetization can happen or not.  
}
\end{abstract}
\maketitle
\section{I. introduction}
The pioneering discovery by Beaurepaire \et\cite{beaure}, that a
femtosecond laser pulse can demagnetize Ni on a subpicosecond time
scale or femtomagnetism, has inspired enormous scientific activities
both experimentally and theoretically \cite{ji, ogasa, zhang, anjan,
  muneaki, bigot}.  The significance of this discovery is that it
demonstrates a possibility for nonthermal writing in a ferromagnetic
medium. One prominent example is the inverse Faraday effect, where the
laser can nonthermally switch spins \cite{stanciu}. This process works
even better when the temperature is lowered \cite{hohlfeld01}.  Its
potential applications require a good understanding of the underlying
excitation mechanism. In spite of extensive investigations in this
field \cite {scholl, hohlfeld00}, how the light transfers the photon
energy to the system and subsequently demagnetizes the sample is still
puzzling though new theoretical and experimental investigations emerge
\cite{bigot}. At the center of the debate is whether there are enough
photons absorbed per atom (estimated at 0.01 \cite{stanciu1}) for magnetic moment change 
\cite{stanciu, stanciu1, koopmans00, koopmans01, koopmans02, dalla, zhang02,
  woodford, wilks01, atxitia, steiauf}. On the one hand, Koopmans \et argue
that the excitation density is too low to induce any substantial change in
magnetization \cite {koopmans00, koopmans01, koopmans02}. If the number of
photons is not enough in the first place, the excitation 
density must be very low. On the other
hand, nearly all the experiments report the magnetization change
either directly or indirectly. A strong demagnetization is
incompatible with a shortage of photons. This puzzle
affects our confidence in femtomagnetism \cite{hertel, vahaplar}.
Therefore, resolving this apparent contradiction is of paramount
importance to femtomagnetism and its future applications.

To this end, there are very few detailed investigations. Koopmans \et
\cite{koopmans04, koopmans02} stated that an effective photon number for
demagnetization can only quench the magnetic moment of 10$^{-4}$
$\rm \mu_{B}$/atom, which is much lower than the observed demagnetization of 0.003 
$\rm \mu_{B}$/atom. In 2007, Dalla \et \cite{dalla} investigated the
influence of photon angular momentum on the ultrafast demagnetization in
Ni. Their results excluded direct transfer of angular momentum to be relevant
for the demagnetization process and showed that the photon contribution to
demagnetization is less than 0.01\%.  
This motivated them and others to search for
alternative mechanisms for strong demagnetization besides the
spin-orbit coupling based mechanism \cite{zhang01}.
Such argument is reiterated by Stanciu \et in
Refs. \cite{stanciu, stanciu1}. Very recently, Hertel \cite{hertel}, in a
viewpoint on Ref. \cite{vahaplar}, claimed that the apparently simple
assumption of a direct transfer of the photon spin to the magnetic system is
not the solution. Majority of research merely avoid this
contradiction by referring to the theoretical argument made 
in the early work \cite{koopmans02} and the circularly polarized experiment in
nickel \cite{dalla}. As we pointed out in a previous paper \cite{zhang02}, the
insensitivity of magnetization change to the light polarization is not a
sufficient condition to rule out the direct involvement of photons in the first place. Laser
affects the spin moment change in two ways. One is that the light
changes the magnetic angular momentum $m_{s}$. This is the case for
the circularly polarized light. The other way is that the light
changes the angular momentum $l$. This is the case for both the
circularly and linearly polarized lights. If the magnetization change
is not sensitive to circularly polarized lights, it only means that
the $ m_{s}$ channel is not effective, but it can not exclude the
$l$-channel. As a result, one can not exclude the photon mechanism. A
very recent theoretical investigation by Woodford \cite{woodford}
reinforces this concept. Nevertheless it is important to note that
irrespective of underlying mechanisms, such a low photon number is
unlikely to induce a substantial magnetization in a sample.

In this paper, we develop a generic scheme to compute the average photons
absorbed per atom and show that for all 24 sets of experimental data
considered here, each atom has more than 1 photon. For a weak laser field,
we examine the relation between the relative demagnetization change and the mean
photons absorbed per atom. Our results show that the small number of photons
absorbed per atom can induce a strong magnetization. Moreover, the linear
dependence of demagnetization change on the amplitude of laser field is
consistent with our perturbation result. For the strong laser intensity, we
resort to the two-level model system. We find that an effective
demagnetization change occurs even with a weak laser field as far as the
system is at resonance.

This paper is structured as follows. Section II presents a formal algorithm to
compute the photon number in femtosecond magnetism. In Sec. III we show the
relation between the relative demagnetization change and the mean photons
absorbed per atom. Then we present a theoretical investigation on the
demagnetization change from the perturbation theory in Sec. IV. For a strong
field, results are presented in a two-level model in Sec. V. Finally, the main
conclusions  of our study are summarized in Sec. VI

\section{II. photon number in femtosecond magnetism}
We start with a laser pulse propagating
along the positive {\it z} axis with the field along the {\it x} axis, 
\be
{\cal E}{_x}(\omega,\tau,z;t)=A_0\exp{(-t^2/{\tau}^2)}\exp{\left
   [i\omega \left (\frac{z}{v}-t\right )\right ]}
\ee
where $A_0$ is the amplitude of laser field, $\tau$ is the pulse
duration, $v$ is the phase velocity, $\omega$ is the laser frequency and $t$ is
the time.
Since $v=c/(n+ik)$, we can rewrite the above equation as
\be
{\cal E}_x(\omega,\tau,z;t)=A_0\exp{(-t^2/{\tau}^2)}\exp{\left
  [i\omega\left (\frac{nz}{c}-t\right )\right ]}\exp{(-\frac{\omega k}{c}z)}
\ee
where $n$ and $k$ are real and imaginary parts of the index of refraction,
respectively, and both are wavelength-dependent. $c$ is the speed of light.

The laser intensity $I$ for the linearly polarized light is computed
from \cite{boyd}
\ba
I(\omega,\tau,z;t)=2n\epsilon_0 c |{\cal E}{_x}(\omega,\tau;t)|^2
=2n\epsilon_0 c A_0^2
\exp{(-2t^2/{\tau}^2)}\exp{(-\frac{2\omega k}{c}z)} \nonumber\\
  \equiv I_{0}(\omega,\tau; t)\exp{(-\frac{z}{d})}
\ea
where $\epsilon_0$ is the permittivity of free space and
$I_{0}(\omega,\tau;t)$ is the laser intensity before penetrating the
sample.  This is the well-known Beer-Lambert law. Here $d$ is the
penetration depth, defined as $d=c/2 \omega k=\lambda /4 \pi k$,
which describes how the light intensity falls off starting from the
surface of a sample. It is clear that $d$ itself is
intensity-independent. Since we are interested in the pulse energy
fluence ${F}(\omega,\tau,z)$, we integrate the above equation over time
and find, \be {F}(\omega,\tau,z) =
2nc\epsilon_0\int^\infty_{-\infty}|{\cal E}{_x}(\omega,\tau,z;t)|^2 dt, \ee
which can be simplified as \be F(\omega,\tau,z) =2nc\epsilon_0
A_0^2\sqrt{\frac{\pi}{2}}\tau \exp{(-\frac{z}{d})}\equiv
F_{max}\exp{(-\frac{z}{d})}.  \ee where $F_{max}$ is the initial laser
fluence given in experiments.  This is the exact expression for the
laser energy fluence at depth $z$, if the pulse is a
Gaussian function.

If a laser pulse of fluence $F(\omega,\tau,z)$ shines on a spot with area $A$,
the total number of photons within $A$ at depth $z$ is \be
N_{photon}(\omega,z)=\frac{F_{max}A}{\hbar\omega}\exp{(-\frac{z}{d})} \ee 
where $\hbar\omega$ is the photon energy, and $\hbar$ is the Planck's constant over $2\pi$.
This equation reveals some crucial information: (i) $N_{photon}$ is a
surface quantity; (ii) the number of atoms illuminated by these
photons must be a surface quantity as well. In other words, we must
compute the photon numbers provided per atom layer by layer, since the light
wave propagates uniaxially.  Consider an fcc structure
with lattice constant $a$ and surface area $A$, the number of atoms in
each layer is \be N_{atom}=\frac{2A}{a^2} \ee where 2 comes from the
fact that each unit cell has two atoms per layer.  The mean number of
photons available to each atom at different depth $z$ is
\begin{equation}
\Xi(z)=\frac{N_{photon}(\omega, z)}{N_{atom}}=\frac{F_{max}}{h\nu}\frac{a^2}{2}\exp{(-\frac{z}{d})}
\equiv \Xi_{max}\exp{(-\frac{z}{d})}.\\
\end{equation}
where $\Xi_{max}$ is the maximum mean number of photons provided to each atom in the top layer.
This equation gives the true mean number of photons provided to each
atom for demagnetization at depth $z$.

Under the low pulse energy fluence, ignoring small reflected photons, which is
justified in highly absorbed metals, we estimate the photons absorbed per atom
at the $j$th layer as
\be
\eta_{j}=\Xi_{max}\left [\exp\left (-\frac{(j-1)a}{2d}\right )-
\exp\left (-\frac{ja}{2d}\right )\right ],\qquad  j=1,2,3,\cdots 
\ee

To appreciate how large $\Xi$ and $\eta_{j}$ are, in Table I we list
the results from 24 different sets of experimental data increasing from top to
bottom. This table is very telling.  
The maximum photon number $\Xi_{max}$ ranges from 1.368 to 90.430. The
photon number at the penetration depth $\Xi (d)$ varies from 0.503 to
33.267. Note that since majority of samples are thinner than the
penetration depth, the number of photons available to each atom exceeds 1.
Here we emphasize that there are enough photons for each atom.
Absorbed photon in the first layer $\eta_{1}$ 
ranges from 0.017 to 1.075. The mean photons absorbed per atom
within the sample is less than $\eta_{1}$. Koopmans' group \cite{koopmans01}
has the smallest value of 0.017. Beaurepaire \et \cite{beaure}'s $\eta_{1}$
value is 0.176, ten times higher than that of Koopmans' $\eta_{1}$. Cheskis \et
\cite{cheskis00} have the largest $\eta_{1}$ value of 1.075 or roughly one
photon per atom, and importantly they already found that the demagnetization
is saturated. {\it This provides a first indication that one does not need a large
amount of photons to demagnetize the sample. The small number of photons
absorbed does not necessarily mean that they can not induce a strong
magnetization change.}

\section{III. demagnetization change vs photons absorbed}
We need to map out the connection between the magnetization change and number of
the photons absorbed. We investigate the relative
demagnetization changes $\Delta M/M$ which are extracted from their
corresponding literatures versus the photons absorbed per atom for a weak
laser intensity \cite{bigot, beaure, cheskis00, atxitia, koopmans00, koopmans01, wilks01}.
We purposely choose a weaker intensity in order to search for a
one-to-one correspondence between the number of photons absorbed and the
amount of magnetization change. For a strong laser, due to the saturation, such
a relation will become complicated. To find some valuable information on the
demagnetization mechanism associated with the number of photons, it's 
necessary to estimate the mean photons absorbed per atom $\bar{\eta}$ more
accurately. Based on Eq. (9), $\bar{\eta}$ is computed from
\be
\bar{\eta}= \frac{1}{k}\sum_{j=1}^{k}\eta_{j}
\ee
where $k$ is the number of layers within the sample's thickness. If the samples'
thickness is thinner than the penetration depth, we calculate $\bar{\eta}$
averaged within the sample's thickness, or else within the penetration
depth. The $\bar{\eta}$ itself includes the reflected and absorbed photons.

Figure 1 shows $\Delta M/M$ as a function of $\bar{\eta}$. Four solid lines
(guide for the eye) from top
to bottom represent the results by  Bigot \et \cite{bigot}, Beaurepaire \et
\cite{beaure}, Cheskis \et \cite{cheskis00} and Atxitia \et \cite{atxitia},
respectively. The corresponding results from Koopmans \et \cite{koopmans00,
  koopmans01} and Wilks \et \cite{wilks01} are also shown in the bottom left
corner of the figure. The vertical dotted line shows that 0.1 photon can
induce about 4\% to 72\% spin momentum change for different laser
durations. This figure is very insightful. (i) The slope of these 
lines approximately characterizes the demagnetization ability of a laser pulse. Each solid line
denotes the results obtained by an identical laser pulse which has the same
pulse duration; (ii) The amplitude of laser $A_{0}$ plays a key role in the
demagnetization change. For the same fluence, as predicted by Eq. (5), the
smaller the pulse duration is, the bigger the laser amplitude becomes. Within the dipole
approximation, the interaction between the laser and the system is dominated
by the amplitude of laser. Therefore, for same $\bar{\eta}$, the amplitudes
increase from bottom to top; (iii) With the same pulse duration parameter, the
demagnetization  change increases along with the photons absorbed per
atom or the laser fluence. The solid lines with slopes of 1.5, 3.8 and
7.4 indicate that {\it one does not need a large amount of photons to induce a
substantial moment change.} For Koopmans \et \cite{koopmans00, koopmans01} and
Wilks \et \cite{wilks01} data, because their laser energy fluences are very
small and the pulse durations are long, the demagnetization changes are much
smaller, but are still consistent with the above picture.

From Fig. 1, it is obvious that the laser amplitude plays a crucial role in
the demagnetization changes. However, is the laser amplitude the only deciding factor for
demagnetization change? The answer is negative. In fact, using the concept of
photons absorbed per atom to explain the strong demagnetization change
obviously neglects two very important factors. First, it doesn't take into
account the interaction between the light and the material. Light has the
energy (how strongly the field oscillates) and the frequency (how fast the
field oscillates with time). $\bar{\eta}$ only takes into account the 
energy but not the frequency, nor the transition matrix elements between
different states. Second, once the number of photons per atom becomes small,
it is well known that the quasi-classical description of photons absorbed per atom for
demagnetization becomes invalid. In particular, when the average number of
photons provided to each atom is less than 40, the electric field behaves
quantum mechanically \cite{3}, i.e., the field oscillates strongly around its
average value. The less the photons are per atom, the stronger the oscillation
is. Therefore, it is necessary to reconsider this hurdle from a different
perspective, and we examine this issue from the perturbation theory and
two-level model, respectively.

\section{IV. perturbation theory and weak intensity}
In Sec. III, it reveals that the relative demagnetization change $\Delta M/M$
is proportional to the mean photons absorbed per atom $\bar{\eta}$ under a
weak laser field. This allows us to treat the laser field perturbatively. 
The Hamiltonian of a system can be described by $H=H^{0}+H^{I}$, where $H^{0}$
is the time-independent Hamiltonian of the unperturbed system,
$H^{I}=-e\vec{\mu}\cdot\vec{\cal E}(t)$ is the time-dependent perturbation. We
start with the Liouville equation $i\hbar\dot{\rho}=[H,\rho]$ for the density
matrix. We keep only the first order term and have  
\be
i\hbar\dot{\rho}^{(1)}=[H^{0},\rho^{(1)}]+[H^{I},\rho^{(0)}].
\ee
where $\rho^{(0)}$ and $\rho^{(1)}$ are the zeroth and first order of the
density matrices, respectively. If we make an unitary transformation as 
$\rho^{(1)}=e^{-iH^{0}t/\hbar} Q e^{iH^{0}t/\hbar}$, and Eq. (11) can be written as 
\be
i\hbar\dot{Q}=e^{iH^{0}t/\hbar}[H^{I},\rho^{(0)}]e^{-iH^{0}t/\hbar}
\ee   
We integrate this equation over time and take $\rho^{(1)}(-\infty)=0$. Then
apply the eigenstate $\langle n|$ on the left and $|m\rangle$
on the right of this equation, and we can obtain 
\be
\rho^{(1)}_{nm}(t)=\frac{1}{i\hbar}(\rho^{(0)}_{nn}-\rho^{(0)}_{mm})
e\mu_{nm}e^{-i\omega_{nm}t-\Gamma t}\int^t_{-\infty}dt'{\cal E}(t')
e^{i\omega_{nm}t'-\Gamma t'}
\ee 
where $\omega_{nm}=(E_{n}-E_{m})/{\hbar}$, $\mu_{nm}$ the transition matrix
elements, and $\Gamma$ the damping factor. Theoretically, the demagnetization
change is involved in the time-dependent density matrix $\rho^{(1)}_{nm}(t)$
through $M_{z}^{(1)}(t)=Tr[S_{z}\rho^{(1)}(t)]$, where $S_{z}$ is the spin
matrix. By investigating the density matrix, we can reveal some crucial
details of the demagnetization. Next we discuss two typical cases. 

\textsl{Case 1}, Continuum wave laser.
Consider a periodical field perturbation $H^{I}=-e\mu A_{0}e^{-i\omega t}$, 
the density matrix becomes
\be
\rho^{1}_{nm}(t)=\frac{e\mu_{nm}}{i\hbar}(\rho^{(0)}_{nn}-\rho^{(0)}_{mm})A_{0}
e^{-2\Gamma t}\frac{(\Delta\omega\sin{\omega t}-\Gamma\cos{\omega t})+
(\Delta\omega\cos{\omega t}+\Gamma\sin{\omega} t)i}{\Delta\omega^2+\Gamma^2}
\ee
where $\Delta\omega=\omega-\omega_{nm}$. Note that 
its frequency dependence is exactly same even if we treat photons quantum
mechanically. It shows that the density matrix $\rho^{(1)}_{nm}(t)$ not only depends on the amplitude
of laser field  $A_{0}$ which is consistent with the conclusion obtained from
Fig. 1, but also depends on the resonance term $(\frac{1}{\Delta \omega^{2}+\Gamma^{2}})$.
Since the spin moment change is directly proportional to $\rho^{(1)}_{nm}(t)$, this predicts
that the magnetic moment change will depend linearly on the amplitude of laser field in the weak
field limit. On the other hand, the resonance term
demonstrates that even for the exactly same laser amplitude, the transition or
the magnetic change can be very different for different frequencies. This has
been largely ignored in the literature.

\textsl{Case 2}, Pulse laser. We assume that laser field has a form
$H^{I}=-e\mu A_{0}e^{-t^{2}/\tau^{2}}e^{-i\omega t}$. Eq. (13) becomes 
\begin{equation}
\begin{aligned}
\rho^{(1)}_{nm}(t)&=\frac{e\mu_{nm}}{i\hbar}(\rho^{(0)}_{nn}-\rho^{(0)}_{mm})
e^{-i\omega_{nm}t-\Gamma t}A_{0}\int^t_{-\infty}dt'e^{-t'^{2}/\tau^{2}}
e^{-i\Delta\omega t'-\Gamma t'}\\
                &\equiv \frac{e\mu_{nm}}{i\hbar}(\rho^{(0)}_{nn}-\rho^{(0)}_{mm})
e^{-i\omega_{nm}t-\Gamma t} R(t)&
\end{aligned} 
\end{equation}
where $R(t)$ is defined as the module of response function. To get some
insightful information about the perturbation of laser field, we choose $\rm \tau=12$ $\rm fs$,
and $ A_{0}=0.05$ V/\AA, and integrate the response function numerically.

Figure 2(a) shows the module of response function $R(t)$ as a function of time
for twelve different energy detuning $\Delta E$ ($\hbar\Delta\omega$) from
0.05  to 0.21 eV. We can see that as time goes by, $R(t)$ first increases to a
peak value, and then settles down to its final value. Importantly the peak
values of $R(t)$ vary a lot for different $\Delta E$ even with the same
laser field amplitude. The biggest one is about 0.9 ($\Delta E$ = 0.05 eV) and
the smallest one is about 0.07 ($\Delta E$ = 0.15 eV). Their difference is
over an order of magnitude. It demonstrates that the transition probability or
spin momentum change can be very large when $\Delta E$ becomes very small or the system
is excited resonantly, even if the field intensity is weak.

Figure 2(b) compares the peak and final values of $R(t)$ as a function of
$\Delta E$. The final $R(t)$ values decrease more quickly with $\Delta E$ than
that of peak ones. The maximum difference is 0.16 at $\Delta E=0.22$ eV, and
the minimum difference is 0.09 at $\Delta E=0.4$ eV. When $\Delta E$ reaches
0.07 eV, the peak and final values are almost same. It implies that regardless
of the amplitude of laser field the transition from one state to another state
is finished when  $\Delta E$ goes to zero. Figure 2(c) shows the time needed
reaching the peak value as a function of $\Delta E$. This indicates that the
peak time is reduced as $\Delta E$ becomes larger. All the above results are
obtained within a weak field limit.  Once the laser field becomes strong, the
first order perturbation becomes invalid. Under this situation, we resort to a
two-level model system.

\section{V. two-level model and strong intensity}
The two-level model has been extensively used in atoms \cite{allen, lesovik},
semiconductors \cite{du, kunikeev, badescu} and
ferromagnetic materials \cite{5}. For a system excited by a laser, the
two-level model can give us a quantitative understanding of the spin
transition. Here we directly quote the Eq. (7) of Ref. \cite{5} in our
previous work about the spin change $\Delta S_{z}$ for a transition from 
state $|a\rangle$ to state $|b\rangle$  
\be
\Delta S_{z}=\aleph_{ab}(t; \omega)(\langle b|s_{z}|b \rangle -\langle a|s_{z}|a \rangle)
\ee
where $t$ is time, and $\omega$ is laser frequency. $\aleph_{ab}(t; \omega)$,
the probability amplitude of finding the system at time $t$ in state $|b
\rangle$, is equal to  
\be
\aleph_{ab}(t;
\omega)=\frac{|W_{ab}|^2}{|W_{ab}|^2+\hbar^2(\omega-\omega_{ba})^2}\sin^2
\left [
  \sqrt{\frac{|W_{ab}|^2}{\hbar^2}+(\omega-\omega_{ba})^2}\frac{t}{2}\right ] 
\ee 
where $W_{ab}=\mu_{ab}A_{0}$. The transition matrix elements $\mu_{ab}$ can be
obtained by calculating the corresponding momentum operator $\vec{p}_{ab}$
from the $ab$ $initio$ calculation \cite{6, 7, 8}.

The probability $\aleph_{ab}(t; \omega)$ is an oscillatory function of time;
for certain values of $t$,  $\aleph_{ab}(t; \omega)=0$, meaning that the system
returns to the initial state $|a\rangle$. At the same time it also
reveals fast oscillation expressed by the latter term of Eq. (17). If 
$|W_{ab}|$ is large enough, the value of $\aleph_{ab}(t; \omega)$ or $\Delta
S_{z}$ can be very large. For the off-resonance excitation, the laser field amplitude
$A_{0}$  controls the demagnetization change entirely. This happens in
semiconductors. {\it For the resonance excitation, or $\omega=\omega_{ba}$,
regardless of how weak the perturbation is, the field can cause the system
to transit from state $|a\rangle$ to state $|b\rangle$ }\cite{10}. Importantly, the
field intensity  only affects the time needed for the system to transit from
$|a\rangle$ to $|b\rangle$, not whether a transition can occur or not. The
smaller the field intensity is, the longer the time it takes.

For a ferromagnetic metal like nickel, the chance that the transition
frequency $\omega_{ba}$ matches that of the laser field is very high
\cite{zhang001}.  This explains the observed demagnetization in those
experiments, in spite of  a relatively weak laser electric
 field.  Figure 3 shows the detailed dependence of the spin change on the
 laser electric field. For a weak  laser field intensity with 0.001 V/\AA, the spin change is
 possible as far as $\omega_{ba}$ matches $\omega$. This demonstrates that the
 strong demagnetization change  is achievable in experiments even if the
 laser field intensity is very weak. If the laser field becomes larger, the
 range of matching frequency becomes broad (see Fig. 3). In semiconductors,
 Pavlov \et \cite{11} showed that increasing temperature reduces the GaAs band
 gap. This may be a test case for our theory.

\section{VI. conclusion}
We have clarified a long-standing conceptual puzzle of the
photon shortage in femtosecond magnetism by comparing the relative
demagnetization change versus the mean photons absorbed per atom. In the weak
laser field, it increases along with the mean photons absorbed per atom
$\bar{\eta}$, which is consistent with the perturbation theory. Importantly,
the results show that few photons absorbed per atom can induce considerable
spin moment changes. The spin moment can be reduced at resonance even if the
field is weak. Our findings overcome a big hurdle in ferromagnetism and should
inspire new experimental and theoretical investigations into the role of
photons and their interactions with electrons in the magnetization change.

\section{acknowledgments}
We would like to thank B. Koopmans (TUE, Netherlands) for valuable communication.
We appreciate Markus M\"unzenberg (G\"ottingen University) for numerous
discussions and for their experimental results before publication of their
preprint Ref. \cite{atxitia}.
This work was supported by the U. S. Department of Energy under Contract
No. DE-FG02-06ER46304, U.S. Army Research Office under Contract
No. W911NF-04-1-0383, the NSFC under No. 10804038, and was also supported by a
Promising Scholars grant from Indiana State University. We acknowledge part of
the work as done on Indiana State University's high-performance
computers. This research used resources of the National Energy Research
Scientific Computing Center at Lawrence Berkeley National Laboratory, which is
supported by the Office of Science of the U.S. Department of Energy under
Contract No. DE-AC02-05CH11231. Initial studies used resources of the Argonne
Leadership Computing Facility at Argonne National Laboratory, which is
supported by the Office of Science of the U.S. Department of Energy under
Contract No. DE-AC02-06CH11357. 

\mbox{$^{*}$gpzhang@indstate.edu}

 \begin{table}
   \caption{The calculated number of photons per atom in
     Ni. $\Xi_{max}$ is the maximum photon number per atom. $d$ is
     the optical penetration depth.  $\Xi(d)$ is the photon number
     at $d$. $\eta_{1}$ is the photons absorbed per atom in the first
     top layer.  The other parameters are taken from the
     literature.  $n$ and $k$ are real and imaginary parts of the index of
     refraction \cite{palik}, respectively. 
     }
   \begin{tabular}{@{}lcccccccl@{}}
\hline
\hline
 No.      &$F$                  &$\lambda$   &$n+ik$         &$d$        &$\Xi_{max}$   &$\Xi(d)$    & $\eta_{1}$  &Ref.     \\
          & ($ \rm mJ/cm^2$)    &  (nm)      &            & (nm)      &                 &               &           &          \\
\hline
1                           &0.6           &729                 &2.28+4.18$i$   &13.93 &1.368 &0.503   &0.017    &\cite{koopmans01}  \\
2                           &0.76      &729
   &2.28+4.18$i$ &13.93 &1.741&0.641   &0.022 &\cite{koopmans00}    \\
3                           &1.4           &790                     &2.46+4.35$i$    &14.45 &3.457&1.272   &0.042     &\cite{wilks01} \\
4                &1.45   &798     &  2.476+4.375$i$ &       14.52&3.617&1.331 &0.044
&\cite{bigot} \\
5                          &1.8           &800                &2.48+4.38$i$   &14.45 &4.501&1.656   &0.055 & \cite{regens00}      \\
6                          &2.04      &729
    &2.28+4.18$i$ &13.93  &4.644&1.708  &0.058&\cite{koopmans00}        \\
7                         &2.0           &785                      &2.45+4.34$i$    &14.50 &4.908&1.805   &0.059 &\cite{dalla}      \\
8                         &2.0           &800                       &2.48+4.38$i$    &14.45 &5.001&1.840   &0.061 & \cite{wilks00}    \\
9                &2.5   &800      &  2.48+4.38$i$ &       14.45 &6.252&2.300 &0.076
&\cite{hohlfeld00} \\
10                         &2.83       &800                       &2.48+4.38$i$   &14.45 &7.075&2.603   &0.086&\cite{melnikov00}       \\
11                &3.5   &800      &  2.48+4.38$i$ &       14.45&8.752&3.220 &0.106
&\cite{hohlfeld00} \\
12                &4.4   &800      &  2.48+4.38$i$ &       14.45&11.003&4.048 &0.133
&\cite{hohlfeld00} \\
13                 &5.3   &800      &  2.48+4.38$i$ &       14.45&13.254&4.876 &0.161
&\cite{hohlfeld00} \\
14                          &7.0           &620
&1.93+3.65$i$    &13.53 &13.566&4.991   &0.176&\cite{beaure, bigot01}      \\
15                         &6.0           &800                         &2.48+4.38$i$   &14.45 &15.004&5.520   &0.182 &\cite{hohlfeld}      \\
16                          &8.0           &620
&1.93+3.65$i$      &13.52 &15.504&5.704 &0.201  &\cite{bigot01} \\
17                          &7.1           &800                 &2.48+4.38$i$   &14.45&17.755&6.532    &0.215 &\cite{regens00}      \\
18                          &12.0           &800                 &2.48+4.38$i$   &14.45 &30.008&11.039   &0.364 &\cite{gudde}      \\
19                          &12.73      &827                &2.53+4.47$i$  &14.73&32.897&12.102    &0.391 &\cite{scholl}    \\
20   &13  &827  &2.53+4.47$i$  &14.73 &33.588&12.356   &0.399 &\cite{rhie}  \\
21                          &13.3          &827               &2.48+4.38$i$   &14.73&34.363&12.642    &0.409 &\cite{cheskis00}      \\
22                          &15.72         &800                     &2.48+4.38$i$   &14.45 &39.308&14.461   &0.476  & \cite{conrad00}     \\
23                          &20            &800                      &2.48+4.38$i$    &14.45  &50.014&18.400  &0.606 & \cite{georg00}        \\
24                          &35            &827                     &2.13+4.17$i$    &14.73   &90.430&33.267 &1.075 & \cite{cheskis00}         \\
\hline
\hline
   \end{tabular}
 \end{table}

 \begin{figure}
\includegraphics[width=15cm]{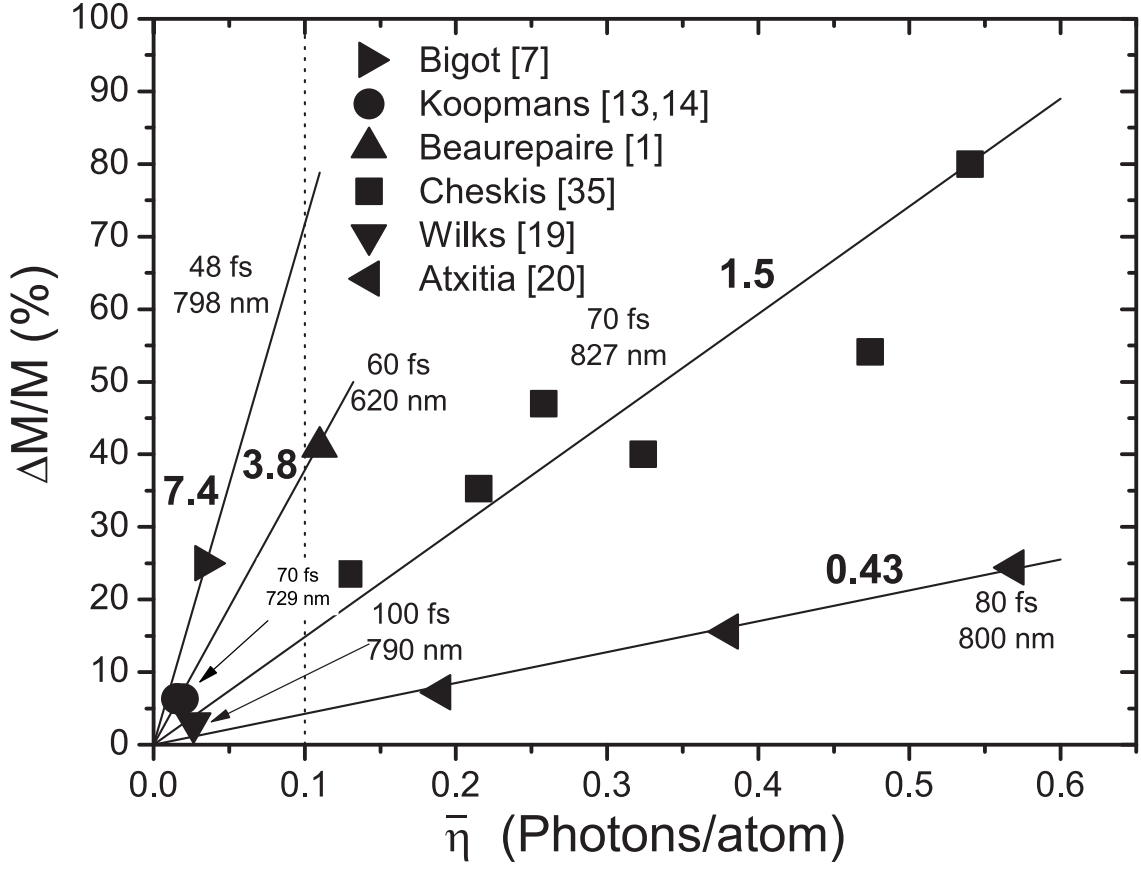}
\caption{Dependence of the relative demagnetization changes on the
         average photons absorbed per atom for six different sets of
         experimental data. The solid lines are guides for the eye,
          whose slopes are shown on the lines. The laser
         pulse durations and the photon wavelengths are also given near
         their data.
\label{Fig1}}
\end{figure}

\begin{figure}
 \includegraphics[width=15cm]{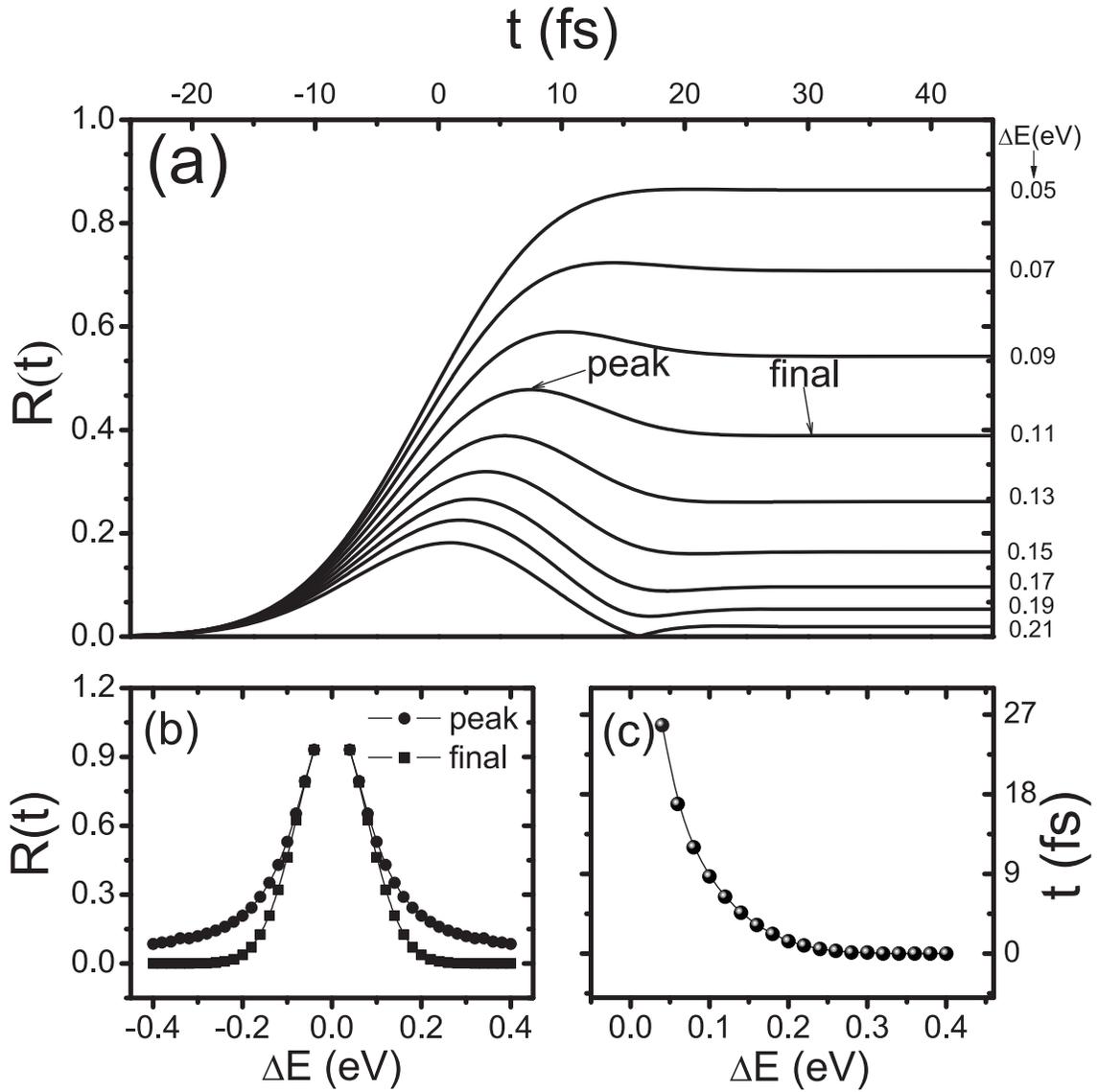}
 \caption{(a) The module of response function $R(t)$ as a function of time at different $\Delta E$.
              (b) The dependence of the peak and final values on $\Delta E$. (c) The time needed
              to reach the peak as a function of $\Delta E$.
\label{Fig2}}
\end{figure}

\begin{figure}
 \includegraphics[width=15cm]{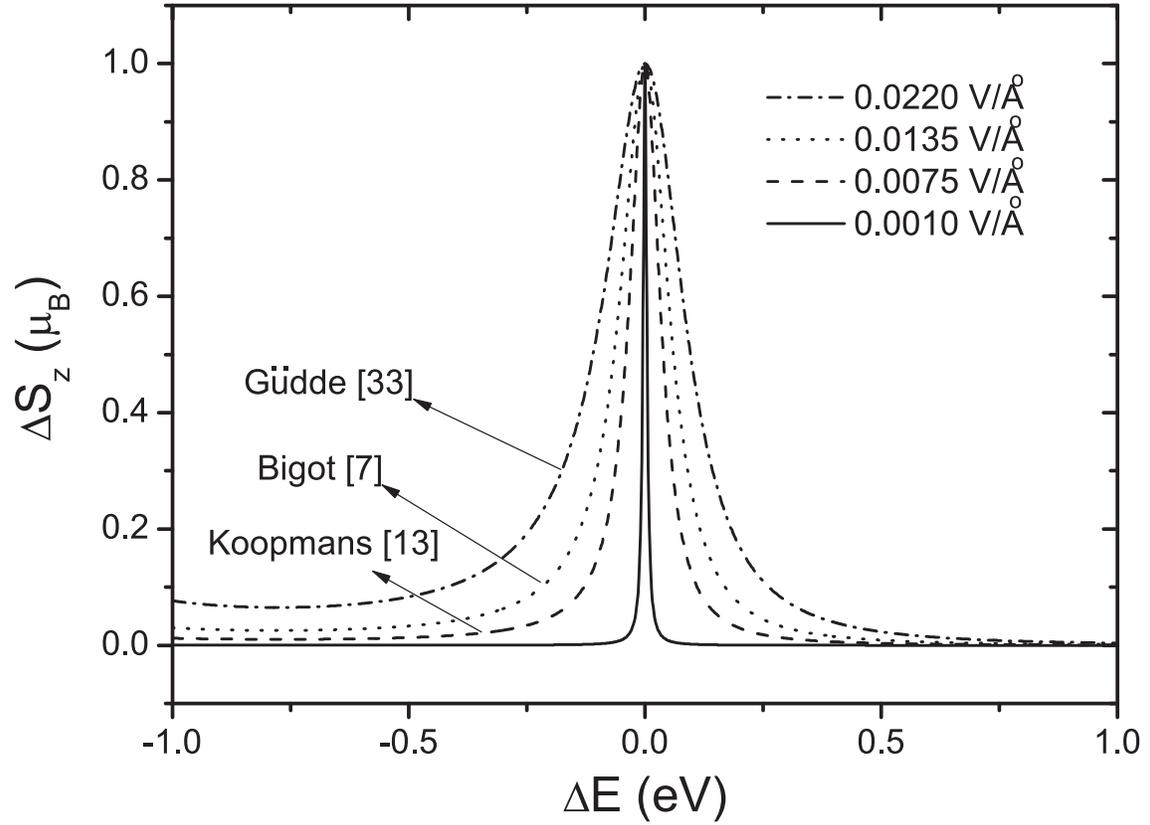}
 \caption{Amplitude of spin momentum change as a function of the detuning $\Delta E=\hbar (\omega -\omega_{ba})$ for 
          three groups at different laser electric fields. The amplitude of laser field is deduced from 
          the experimental laser fluence based on the Eq. (5). All the curves are shifted to zero at resonance.
\label{Fig3}}
\end{figure}

\end{document}